\documentclass[twocolumn,showpacs,preprintnumbers,amsmath,amssymb]{revtex4}

\usepackage{graphicx}
\usepackage{dcolumn}
\usepackage{bm}

\begin{document}

\preprint{APS/123-QED}

\title{Classical rotational inertia of solid $^4$He}

\author{J.G. Dash$^{1}$ and J.S. Wettlaufer$^{2}$}
\affiliation{$^{1}$University of Washington, Seattle, Washington, 98195-1560}
\affiliation{$^{2}$Yale University, New Haven, Connecticut, 06520-8109}

\date{\today}

\begin{abstract}
The observation of reduced rotational inertia in a cell 
containing solid $^4$He has been
interpreted  as evidence for superfluidity of the solid. An 
alternative explanation is slippage of the solid
at the container wall due to grain boundary premelting between the 
solid and dense adsorbed layers at the
container  wall. We calculate the range of film thickness and the 
viscous drag, and find that the
slippage can account for the observations.

\end{abstract}

\maketitle

A recent report by Kim and Chan \cite{Chan1} describes the observation of 
inertial anomalies, termed ``non
classical rotational inertial fraction (NCRIF)''  in solid
$^4$He, which are taken to  demonstrate superfluidity of the solid. 
The exciting possibilities
raised by the experiment and their earlier study \cite{Chan2} wherein $^4$He 
was solidified
in {\it Vycor}, a  microporous glass, span a range of fundamental 
issues in quantum materials.  Leggett
notes \cite{Leggett} the possibility of supersolid behavior  as a long standing 
speculation, based on the
hypothesis of  Bose-Einstein condensation of zero-point vacancies, 
and Prokof'ev and
Svistunov argue \cite{Prokofev} that the vacancy density in well ordered solid $^4$He is
insufficient for Bose-Einstein condensation at the experimental 
temperatures, speculating instead
that the NCRIF in the Vycor study \cite{Chan2} may be due to the 
superfluidity of vacancies in a
defect-laden layer of solid $^4$He at the Vycor surfaces.  Finally,  Beamish \cite{Beamish}  noted the possibility that a disordered thin liquid like layer at the Vycor walls may persist at low temperature and argued that its superfluid properties would be different from those found at pressures well below solidification pressure.

Here we suggest an alternative explanation of the experiment on bulk solid [1]: slippage of the solid
at the wall of the container due to a liquid film caused by grain boundary premelting. The premelting
in this case is not at ordinary grain boundaries, but at the interface  between the bulk solid and dense adsorbed layers at the container  wall. The dense layers, due to strong 
adsorption forces, are responsible for nonzero wetting angles between solid $^4$He and 
copper and glass walls \cite{Balibar, Dash82}; rather than an ordinary grain  boundary, the contacting surface in 
question more nearly resembles
the interface between two different materials. The liquid film 
separates the bulk solid from the
torsional balance and replaces the shear strength of the dry 
interface by viscous drag. Our model
allows the possibility that the film is superfluid, in which case the 
calculation is equally
applicable to its normal fluid component. In the following we first 
estimate the thickness of the
premelted film, and then calculate the dynamical coupling.

The nature of premelting in any system is determined by the 
competition between bulk and surface free
energies. {\it  Complete premelting}, in which the thickness of the 
melt layer diverges as temperature
approaches the bulk transition, requires that the total excess 
surface free energy per unit area,
$F(d)$, be a positive  monotonically decreasing function of the film 
thickness with a global minimum at
infinite film thickness. In {\it incomplete premelting}, the melt 
thickness remains finite at the bulk
transition, such as in recent studies of ice \cite{ice}. It is a general 
result \cite{schick}-\cite{Benatov} that in a symmetrical
system (e.g., solid/liquid/solid) the long range interactions are 
attractive, and consequently grain
boundary melting must be incomplete.  The most complete formulation 
of the excess
surface free energy per unit area, $F(d)$, for systems {\it entirely}
controlled by frequency dependent
dispersion forces is that of Dyzaloshinskii,  Lifshitz and Pitaevskii 
(DLP) \cite{DLP}. However, it requires
as input the frequency dependent  dielectric properties of the layers 
in the system under
consideration.  Because we are dealing with bulk solid $^4$He and the 
dense adsorbed solid at the
container wall, the input data for the DLP theory are not available. 
Therefore, we proceed
with reasonable ranges of the Lennard-Jones parameters \cite{Cole, vdws}.

The total free energy of the system at a given temperature $T$ and
pressure $P$ is written
\begin{equation}
G_T(T,P,d) =\rho_{\ell}\mu_{\ell}(T,P)d + F(d), 
\end{equation}
where the liquid density and chemical potential are $\rho_{\ell}$ and $\mu_{\ell}(T,P)$.
In grain boundary premelting
\begin{equation}
F(d) \equiv
\gamma_{ss}(d)= 2\gamma_{s{\ell}} + \rho_{\ell}{\int_{d}^{\infty}} V(z) dz,
\end{equation}
where $\gamma_{ss}(d)$ is the interfacial free energy of
the solid--solid interface, and  $\gamma_{s{\ell}}$ is the
interfacial free energy per unit area of the solid--liquid
interfaces, with implicit reference to the crystallographic
orientation present at an interface. In lieu of the DLP theory, the
most general phenomenological mean field model considers $V(z)$ as
the Lennard-Jones potential \cite{Cole} but  augmented to embody the effects
of retardation viz a viz
\begin{equation}
V(z) = {\frac{4 {C_3}^3}{27
D^2}}{\frac{1}{z^9}} - {\frac{C_3}{z^3}} - {\frac{B}{z^4}},
\end{equation}
where $z$ is the coordinate normal to the surfaces, ${C_3}$ ($B$) is the
nonretarded (retarded) van der Waals attraction, and $D$ is the well
depth.  At each temperature and pressure below $T_\lambda$ the bulk
and interfacial free energies strike a balance and one can show
[e.g., \cite{Lipowsky, RPP} that a unique equilibrium film thickness obtains from
\begin{equation}
\frac{1}{\rho_{\ell}} \frac{\partial F(d) }{\partial d} = - q_m { \frac{T_\lambda - T}{T}} \equiv - q_m t,
\end{equation}
where the latent heat of fusion is $q_m$ and $t$ is the reduced temperature.  From
this we can simply write the equilibrium film thickness--temperature relation as $t = {q_m}^{-1} V(d)$.

In the figure we plot $d = d(t)$ for a range of potential parameters suggested from the detailed
analysis of the wetting of wide classes of substrates by liquid
helium \cite{Cole}.  Curves for three values of the latent heat of fusion
$q_m$ are shown and the value at the NCRIF onset temperature of 175 mK
is extrapolated to lower temperatures from 1 K using the data of
Swenson \cite{Swenson}.  Because $t = {q_m}^{-1} V(d)$,  we find $q_m$ to be
the most important parameter in the problem: The latent heat embodies
the bulk free energy penalty for converting solid to liquid,
against the melt driving interactions of the potential.  Therefore,
the general understanding that $q_m \rightarrow 0$ as $T \rightarrow
0$ ($t \rightarrow 1$), but the lack of experimental information on
$q_m$, at solidification pressures and temperatures below about 1K,
leaves open important quantitative questions.  Thus, we view our film
thickness calculations as conservative--thinner than is likely.  We find that although the
magnitude of the thickness of a premelted layer depends on the
parameters used in the calculation, (a) the temperature dependence
itself is rather weak and (b) the film thicknesses (1-4 atomic
layers) are sufficient to accommodate flow and superflow (e.g., \cite{Girvin, Hallock}).

We now estimate the dynamical coupling of the solid to the container,
through the premelted layer. The solid is a thin walled hollow cylinder, of mean radius $a$, wall
thickness $s$ and height $h$, bathed on both sides by a layer of liquid thickness $d$.  The liquid is driven by the container's torsional oscillations of angular displacement  $\theta(t)=\theta_o e^{i\omega t}$, driving the solid at the same frequency, and angular amplitude 
$\theta'_o$. The hydrodynamic regime is governed by the relative magnitudes of $d$ and the decay length of transverse viscous waves $\lambda=\sqrt{2 \eta/ \rho_{\ell} \omega}$, where the dynamic viscosity is $\eta$ and the density of the liquid $\rho_{\ell}$ \cite{Lamb, DashTaylor}.  To evaluate $\lambda$ we 
have the experimental frequency \cite{Chan1} $\omega = 2.3\times 10^3 s^{-1}$ and the liquid density \cite{Swenson} $\rho_{\ell} \approx 0.2 g/cm^3$ in the range of experimental pressures. Since the viscosity of the premelted liquid 
is not known, we must estimate it from the properties of the bulk liquid. The viscosity of normal 
liquid HeI is about $20 \mu P$; the viscosity of the normal component of HeII between 1.2 K and 
$T_{\lambda}$ ranges between $10 \mu P$ and $20 \mu P$ \cite{DashTaylor}. However, the viscosity may be enhanced in the very narrow gap $d$ \cite{Hallock}, thereby placing an uncertainty on the experimental value of $\eta$.  As a conservative measure, we will allow a 
possible enhancement of an order of magnitude.  Accordingly, we estimate $2\times 10^{-4}<\lambda< 
7\times 10^{-4}$ cm, which is on the order of one thousand times the thickness $d$ of the grain 
boundary melted liquid, thereby reducing the problem to one of slow, nearly steady flow.  A second 
simplification stems from the ratio of $d$ to the cylinder radius $a$; $d/a<<1$, which makes the problem 
equivalent to the drag between parallel plates. Therefore the fluid velocity varies linearly 
between the surface of the cell and the solid helium, so that the viscous drag per unit area 
on the solid helium is $f = (a\omega\eta/d)(\theta_o-\theta'_o) e^{i\omega t}$. The torque ${\bf{\cal T}}_T$ 
due to the total force on both inner
and outer surfaces of the solid cylinder is
\begin{equation}
{\bf{\cal T}}_T = {{4\pi a^3 h\omega\eta}\over{d} } (\theta_o-\theta'_o)e^{i\omega t}.
\end{equation}
The torque induces the solid's inertial
response, the time rate of change of angular momentum: ${\dot{\bf{L}}}_s = I\omega^2\theta'_oe^{i\omega t}$, where $I=2\pi a^3 hs
\rho_{sol}$, in which we emphasize that  $\rho_{sol}$ is the density of the solid to avoid confusion with the conventional nomenclature for the superfluid density.  Thus, the fractional difference in the amplitudes of rotational motion between the bulk solid and the premelted film is written as $\frac{\theta_o-\theta'_o}{\theta'_o} \equiv \delta $, and hence the fractional inertial response is $\frac{\theta_o-\theta'_o}{\theta_o} = \frac{\delta}{1+ \delta}$, wherein the controlling parameter is the value of $\delta$ which is written as follows
\begin{equation}
\delta = {{ds }\over{\lambda^2}}  {{\rho_{sol}}\over {\rho_{\ell}}},
\end{equation}
The ranges of possible values of $d$ and $\eta$
impose wide limits on the estimate of slippage:  $0.004<\delta<0.26$. 
These extremes span the maximum
values of the NCRIF measured at low temperature and low amplitude, 
roughly 0.005 to 0.02, depending on the
pressure \cite{Chan1}.

Additional observations of Kim and Chan are that 
the NCRIF decreases at higher
amplitudes of oscillation, which is given as strong supporting 
evidence of superfluid solid behavior,
attributing it to exceeding the critical velocity of superfluidity in 
the solid \cite{Chan1}. We propose that it
can indeed be evidence of superfluidity, but in the liquid film 
rather than in the solid. The slippage
is controlled, at low relative speeds, by the viscosity of the normal 
fluid component, and is augmented,
at higher speed, by the excitation of the superfluid fraction.

The results of the earlier work with Vycor \cite{Chan2} are relevant here.  Because of pinning of the solid by the Vycor matrix \cite{Chan2}, the loss of rotational inertia is interpreted as an indication of
superfluidity. It may be, as Prokof'ev and Svistunov \cite{Prokofev} and Beamish \cite{Beamish} suggest, in 
a disordered solid layer at the
Vycor interface. But it can be expected that grain boundary melting 
occurs in the Vycor study, as in the
experiment with bulk solid. Thus the superfluidity would be 
associated with the liquid film, rather than
the solid.  Indeed, the nm scale disorder of Vycor was used in order to enhance the density of vacancies \cite{Chan2}, and depending on pressure and the assumed fraction of material locked by the tortuosity of the Vycor, the estimate of the BEC fraction could be  {\it less} than that in the bulk solid.   In the framework of our theory, the difference between the experiments may be explained in terms of the wall materials inducing a different density of adsorbed solid.   Interesting tests that span the existing range of behavior would involve the insertion of a ridge in the cell wall of the most recent experiment \cite{Chan1} or changing the wall material.   An appropriate ridge dimension would lock the solid and thereby offer a possible test of whether slippage had explained the NCRIF.  An additional observation is an extreme sensitivity of the NCRIF to $^3$He; in our model it is explained by the very small fraction of the 
sample that is involved in the grain boundary melt, and the greater 
solubility of $^3$He in liquid, rather than in solid helium.

Finally, although we propose an alternative explanation to a superfluid solid, we consider that decoupling due to a premelted, and possibly superfluid, film offers interesting possibilities. The film exists in a region of pressure and temperature not otherwise accessible, and the experiment suggests interesting and entirely new possibilities in studies of premelting and
liquid confinement.

\begin{acknowledgments}
We would like to acknowledge conversations with S.M. Girvin and the support of the National Science Foundation, the Bosack and Kruger Foundation and Yale University. 
\end{acknowledgments}

\begin{figure}[h] 
	\begin{center}
		\includegraphics[width=.30\textwidth]{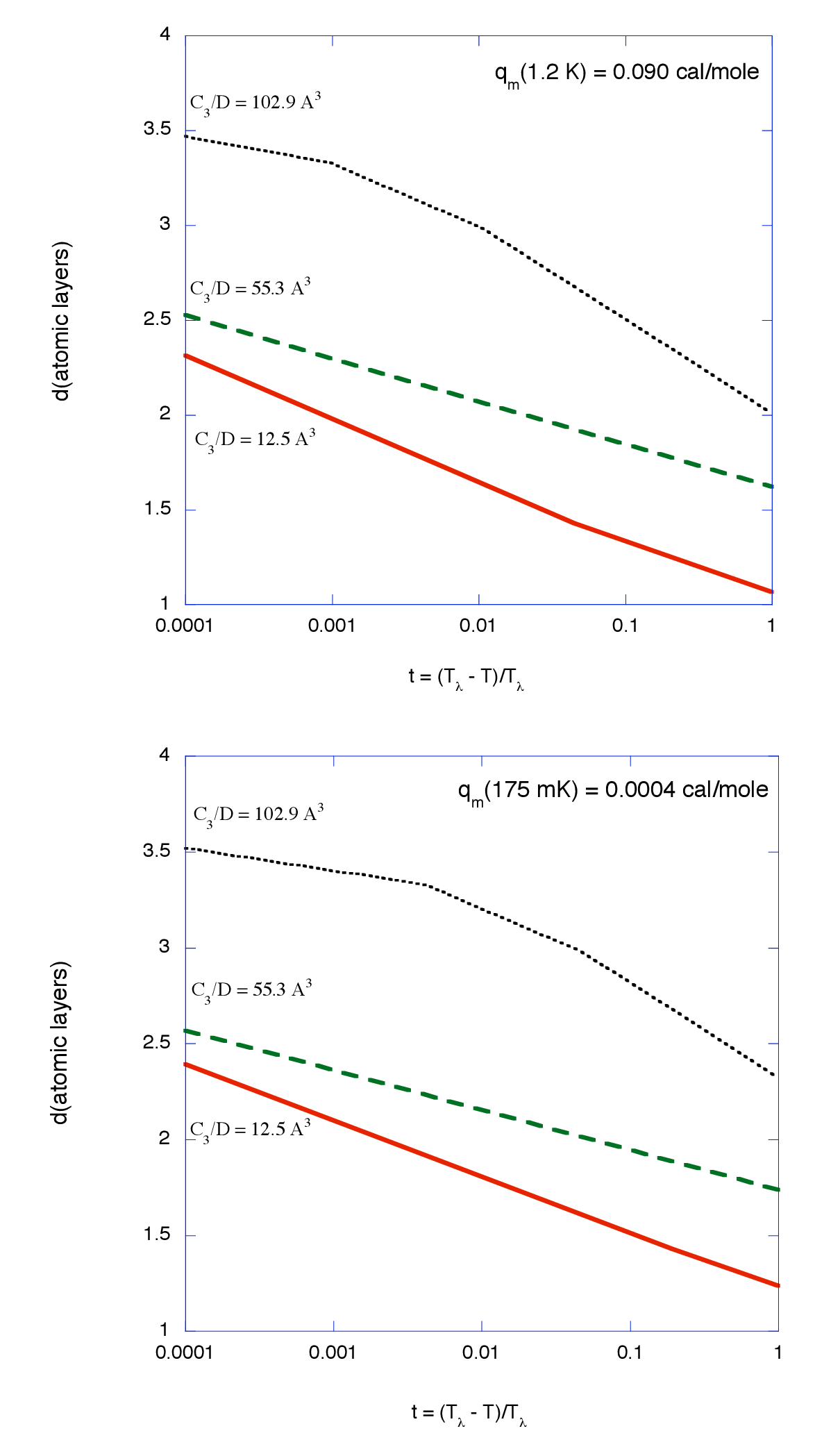}
	\end{center}
	\caption{\label{one} Plots of liquid $^4$He
film thickness, in units of atomic layers, as a function of reduced temperature, $t$, and values of the Lennard-Jones potential parameters ${C_3}/D$ =  102.9, 55.3,  and 12.5 ${\AA}^3$ (as labeled adjacent to the curves) chosen for expected ranges \cite{Cole} and because
retardation typically occurs at ranges that are much larger than is
found, we let $B$ =0.   The two different sets of three curves are
meant to demonstrate the important influence of the latent heat of
fusion, $q_m$; (a) $q_m(T = 1.2K)$ = 0.09 cal/mole, and (b) $q_m(T = 175 mK)$ = 0.0004 cal/mole.
The data from Swenson \cite{Swenson} are extrapolated to the experimental
temperature of 175 mK based on the requirement that $q_m$ vanishes at
absolute zero.}
\end{figure}

\end{document}